\documentclass[12pt]{article}%
\usepackage{amsmath}
\usepackage{amsfonts}
\usepackage{amssymb}
\usepackage{graphicx}%
\providecommand{\U}[1]{\protect\rule{.1in}{.1in}}
\newcommand{\be}{\begin{equation}}
\newcommand{\ee}{\end{equation}}
\newcommand{\ben}{\begin{equation*}}
\newcommand{\een}{\end{equation*}}
\newcommand{\ar}{\begin{array}}
\newcommand{\arn}{\end{array}}

\def\pnot{\mbox{${\not{\hbox{\kern-3.0pt$p$}}}$}}
\def\qnot{\mbox{${\not{\hbox{\kern-2.0pt$q$}}}$}}
\def\enot{\mbox{${\not{\hbox{\kern-2.0pt$e$}}}$}}
\def\knot{\mbox{${\not{\hbox{\kern-2.0pt$k$}}}$}}

\def\fun#1#2{\lower3.6pt\vbox{\baselineskip0pt\lineskip.9pt\ialign
{$\mathsurround=0pt#1\hfil##\hfil$\crcr#2\crcr\sim\crcr}}}

\begin{document}
\begin{titlepage}
\hfill Budker INP 2011-27
\vspace{0.7cm}

\begin{center}
{\bf Connection between complete and M\"obius forms of gauge invariant
operators $^{\ast}$}
\end{center}
\vskip 0.5cm
\vskip 0.5cm \centerline{V.S.~Fadin$^{a\,\dag}$,
R.~Fiore$^{b\,\ddag}$, A.V.~Grabovsky$^{a\,\dag\dag}$, 
A.~Papa$^{b\,\ddag\dag}$} \vskip .6cm
\centerline{\sl $^{a}$ Budker Institute of Nuclear Physics, 630090
Novosibirsk, Russia} 
\centerline{\sl and Novosibirsk State University,
630090 Novosibirsk, Russia} \centerline{\sl $^{b}$ Dipartimento di
Fisica, Universit\`a della Calabria,} 
\centerline{\sl and Istituto
Nazionale di Fisica Nucleare, Gruppo collegato di Cosenza,}
\centerline{\sl Arcavacata di Rende, I-87036 Cosenza, Italy}
\vskip 1cm
\begin{abstract}
We study the connection between complete representations of gau\-ge invariant
operators and their M\"{o}bius representations acting in a limited space of
functions. The possibility to restore the complete representations from
M\"{o}bius forms in the coordinate space is proven and a method of
restoration is worked out. The operators for transition from the standard
BFKL kernel to the quasi-conformal one are found both in M\"{o}bius and
total representations.
\end{abstract}
\vskip 1cm
\hrule \vskip.3cm \noindent $^{\ast}${\it Work supported
in part  by  grant  14.740.11.0082 of Federal Program ``Personnel of
Innovational Russia'', in part by  RFBR grant 10-02-01238, in part by Dynasty
foundation, and in part by Ministero Italiano dell'Istruzione,
dell'Universit\`a e della Ricerca.} \vfill \vfill

\vspace{0.5cm}
\vfill

$
\begin{array}{ll} ^{\dag}\mbox{{\it e-mail address:}} &
\mbox{FADIN@INP.NSK.SU}\\
^{\ddag}\mbox{{\it e-mail address:}} &
\mbox{FIORE@CS.INFN.IT}\\
^{\dag\dag}\mbox{{\it e-mail address:}} &
\mbox{A.V.GRABOVSKY@INP.NSK.SU}\\
^{\ddag\dag}\mbox{{\it e-mail address:}} &
\mbox{PAPA@CS.INFN.IT}\\
\end{array}
$
\end{titlepage}

\vfill \eject

\section{Introduction}

\label{sec:introduction} The BFKL (Balitsky-Fadin-Kuraev-Lipatov)
approach~\cite{BFKL} was formulated in momentum space. In this space the
kernel of the BFKL equation was calculated in the next-to-leading order (NLO)
long ago, at first for forward scattering (i.e. for $t=0$ and color singlet in
the $t$-channel)~\cite{Fadin:1998py} and then for any fixed (not growing with
energy) momentum transfer $t$ and any possible two-gluon color state in the
$t$-channel~\cite{FF05}. Unfortunately, the NLO kernel is rather complicated.
In the colour singlet case at $t\neq0$ it consists of numerous intricate
two-dimensional integrals.

In the case, most interesting for phenomenological applications, of colourless
particles scattering, the leading order (LO) BFKL kernel has a remarkable
property~\cite{Lipatov:1985uk}. In the M\"{o}bius representation, i.e. in the space
of functions vanishing at coinciding transverse coordinates (impact parameters)
of Reggeons, it is invariant with regard to conformal transformations of these
coordinates. Moreover, in this representation the kernel
coincides~\cite{Fadin:2006ha} with the kernel of the colour dipole
model~\cite{dipole} formulated in the impact parameter space.

In the NLO one could expect that in the M\"{o}bius representation the BFKL kernel
would be quasi-conformal, i.e. its conformal invariance would be violated only
by terms proportional to $\beta$-function, so that it would remain unbroken in
$N$=4 supersymmetric Yang-Mills theory ($N$=4 SUSY). However, the direct
transformation of the known QCD kernel from the momentum into the impact
parameter space, with subsequent transition into the M\"{o}bius representation,
gives a kernel which is not
quasi-conformal~\cite{Fadin:2006ha,Fadin:2007ee,Fadin:2007de}.
In $N$=4 SUSY the conformal invariance of
the kernel obtained in such a way is also broken~\cite{Fadin:2007xy}. Then,
the kernel of the colour dipole model was calculated in the NLO directly in
the impact parameter space~\cite{Kovchegov:2006vj,Balitsky:2008zz}. It
turned out that the result differs from the one obtained by transformation
from the momentum space.

However there is an ambiguity in the definition of low-$x$ evolution kernels at
NLO. It is analogous to the well-known ambiguity of NLO anomalous dimensions
and follows from the possibility to redistribute radiative corrections
between kernels and impact factors. This ambiguity was discussed in detail
in Ref.~\cite{Fadin:2009za}. It has been shown recently~\cite{Fadin:2009gh}
that it permits one both to reach agreement with the colour dipole model (with
account of the improvement of the result in Ref.~\cite{Balitsky:2008zz} made
in Ref.~\cite{Balitsky:2009xg}) and to obtain quasi-conformal shape for the
kernel in the M\"{o}bius representation. This shape appears to be quite simple.
It is unbelievably simple in comparison with the known shape of the kernel in
the momentum space. We will call ``standard'' the kernel and the impact
factors defined in the NLO in Ref.~\cite{Fadin:1998fv} in the space of
transverse momenta of two interacting Reggeons. Evidently, the question arises
about the relation between these two shapes.

Transition from the ``complete'' to the  M\"{o}bius representation means
restriction of the complete space of functions where the kernel is defined to
the space of functions which are equal to zero at coinciding values of
Reggeon impact parameters. Therefore, the interrelation between the latter
representation and the complete one is not obvious. In
particular, the possibility to restore the complete representation from the
M\"{o}bius one is questionable. Our paper is devoted to the discussion of this
problem. It is organized as follows. In the next Section all necessary
notations and definitions are given. In Section~\ref{sec:interrelation}
the equivalence of complete and M\"{o}bius representations for gauge invariant
operators is proven. In Section~\ref{sec:restoration} the complete
representation of the operator connecting standard and quasi-conformal
kernels is restored from its M\"{o}bius representation. In
Section~\ref{sec:mobius} the interrelation of complete and M\"{o}bius
representations is illustrated on the example of this operator. Conclusions
are drawn in Section~\ref{sec:conclusion}.

\section{ Notation and definitions}
\label{sec:notation}

For brevity we will use, as in Ref.~\cite{Fadin:2006ha},
states $|\vec{q}\rangle$ with definite two-dimensional (we will not use
dimensional regularization and put the space-time dimension $D=4$) transverse
Reggeon momentum $\vec{q}$ and states $|\vec {r}\rangle$ with definite
Reggeon impact parameter $\vec{r}$  normalized as follows:
\begin{equation}
\langle\vec{q}|\vec{q}^{\;\prime}\rangle=\delta(\vec{q}-\vec{q}^{\;\prime
})\;,\;\;\;\;\;\langle\vec{r}|\vec{r}^{\;\prime}\rangle=\delta(\vec{r}-\vec
{r}^{\;\prime})\;,\;\;\;\;\;
\langle\vec{r}|\vec{q}\rangle=\frac{e^{i\vec{q}\,\vec
{r}}}{2\pi}\;.\label{normalization}%
\end{equation}
The kernel defined in Ref.~\cite{Fadin:1998fv} is represented by an
operator $\hat{{K}}$. It is given by the sum of virtual (related to the gluon
Regge trajectory) and real (related to real particle production in Reggeon
collisions) parts, so that
\begin{equation}
\hat{{K}}=\hat{\omega}_{1}+\hat{\omega}_{2}+\hat{{K}}_{r}%
\;.\label{operator of the BFKL kernel}%
\end{equation}
Here $1$ and $2$ are Reggeon  indices,
\begin{equation}
\langle\vec{q}_{i}|\hat{\omega}_{i}|\vec{q}_{i}^{\;\prime}\rangle=\delta
(\vec{q}_{i}-\vec{q}_{i}^{\;\prime})\omega(-\vec{q}_{i}^{\;2}%
)\;,\label{trajectory ff}%
\end{equation}
$\omega(t)$ is called the gluon trajectory (although, in fact, the
trajectory is $1+\omega(t)$), and
\begin{equation}
\langle\vec{q}_{1},\vec{q}_{2}|\hat{{K}}_{r}|\vec{q}_{1}^{\;\prime},\vec
{q}_{2}^{\;\prime}\rangle=\delta(\vec{q}_{1}+\vec{q}_{2}-\vec{q}_{1}%
^{\;\prime}-\vec{q}_{2}^{\;\prime})\frac{1}{\sqrt{\vec{q}_{1}^{\,\,2}\vec
{q}_{2}^{\,\,2}}}{K}_{r}(\vec{q}_{1},\vec{q}_{1}^{\;\prime};\vec{q})\frac
{1}{\sqrt{\vec{q}_{1}^{\,\,\prime2}\vec{q}_{2}^{\,\,\prime2}}}%
\;,\label{real kernel}%
\end{equation}
where $\vec{q}=\vec{q}_{1}+\vec{q}_{2}$ and ${K}_{r}(\vec{q}_{1},\vec{q}%
_{1}^{\;\prime};\vec{q})$ is defined in Ref.~\cite{Fadin:1998fv}. The
appearance of square roots can seem strange to an experienced reader; these
square roots are connected with the normalization~(\ref{normalization}) and
can be removed by passing to states $|\vec{q}\rangle$ normalized as
$\langle\vec{q}|\vec {q}^{\;\prime}\rangle=\vec{q}^{\;2}\delta(\vec{q}
-\vec{q}^{\;\prime})$. In the LO
\begin{equation}
{K}_{r}^{(B)}(\vec{q}_{1},\vec{q}_{1}^{\;\prime};\vec{q})=\frac{\alpha_{s}N_c}
{2\pi^{2}}\left(  \frac{\vec{q}_{1}^{\;2}\vec{q}_{2}^{\;\prime\;2}+\vec
{q}_{1}^{\;\prime\;2}\vec{q}_{2}^{\;2}}{\vec{k}^{\;2}}-\vec{q}^{\;2}\right)
~,\label{real lo kernel}%
\end{equation}
where $\vec{k}=\vec{q}_{1}-\vec{q}_{1}^{\;\prime}=\vec{q}_{2}^{\;\prime}%
-\vec{q}_{2}$ and the superscript ${(B)}$ denotes leading order.

In terms of the kernel $\hat{{K}}$, the $s$-channel discontinuities of
scattering amplitudes for processes $A+B\rightarrow A^{\prime}+B^{\prime}$
are presented as
\begin{equation}
-4i(2\pi)^{2}\delta(\vec{q}_{A}-\vec{q}_{B})\mbox{disc}_{s}\mathcal{A}%
_{AB}^{A^{\prime}B^{\prime}}=\langle A^{\prime}\bar{A}|\left(  \hat{\vec{q}%
}_{1}^{\;2}\hat{\vec{q}}_{2}^{\;2}\right)  ^{-\frac{1}{2}}e^{Y\hat{{K}}%
}\left(  \hat{\vec{q}}_{1}^{\;2}\hat{\vec{q}}_{2}^{\;2}\right)  ^{-\frac{1}%
{2}}|\bar{B}^{\prime}B\rangle\;, \label{discontinuity representation}%
\end{equation}
where  $Y=\ln(s/s_{0})$, $s_{0}$ is an
appropriate energy scale,$\;q_{A}=p_{A^{\prime}}-p_{A},\;q_{B}
=p_{B}-p_{B^{\prime}}$. The states $\langle
A^{\prime}\bar{A}|$ and $|\bar{B}^{\prime}B\rangle$ are normalized in such a
way that
\begin{equation}
\langle\vec{q}_{1},\vec{q}_{2}|\bar{B}^{\prime}B\rangle=4p_{B}^{-}\delta
(\vec{q}_{B}-\vec{q}_{1}-\vec{q}_{2}){{\Phi}_{B^{\prime}B}(\vec{q}_{1},\vec
{q}_{2})}\;, \label{impact BB}%
\end{equation}%
\begin{equation}
\langle A^{\prime}\bar{A}|\vec{q}_{1},\vec{q}_{2}\rangle=4p_{A}^{+}\delta
(\vec{q}_{A}-\vec{q}_{1}-\vec{q}_{2}){{\Phi}_{A^{\prime}A}(\vec{q}_{1},\vec
{q}_{2})}\;, \label{impact AA}%
\end{equation}
with $p^{\pm}=(p_{0}\pm p_{z})/\sqrt{2}$ and the impact factors $\Phi$
expressed through the Reggeon vertices according to Ref.~\cite{Fadin:1998fv}.

The kernel $\hat{K}$ is symmetric, as it can be seen from Eqs.~(\ref{operator of the BFKL
kernel})--(\ref{real lo kernel}), i.e. $\hat{{K}}=\hat{{K}}^{T}$ or
\begin{equation}
\langle\vec{q}_{1},\vec{q}_{2}|\hat{{K}}|\vec{q}_{1}^{\;\prime},\vec{q}%
_{2}^{\;\prime}\rangle=\langle\vec{q}_{1}^{\;\prime},\vec{q}_{2}^{\;\prime
}|\hat{{K}}|\vec{q}_{1},\vec{q}_{2}\rangle~.
\end{equation}
However, the kernel which is conformally invariant in the M\"{o}bius
representation in the LO~\cite{Lipatov:1985uk, Bartels:2004ef} is not $\hat
{K}$, but the non-symmetric kernel
\begin{equation}
\hat{\mathcal{K}}=\left(  \hat{\vec{q}}_{1}^{\;2}\hat{\vec{q}}_{2}%
^{\;2}\right)  ^{-\frac{1}{2}}~\hat{{K}}\left(  \hat{\vec{q}}_{1}^{\;2}%
\hat{\vec{q}}_{2}^{\;2}\right)  ^{\frac{1}{2}}. \label{cal K}%
\end{equation}
The transition to this kernel is possible thanks to the invariance of the
discontinuity~(\ref{discontinuity representation}) with respect to the
transformation
\[
\hat{{K}}\rightarrow\hat{\mathcal{O}}^{-1}\hat{{K}}\hat{\mathcal{O}
}~,\;\;\;\;\;
\langle A^{\prime}\bar{A}|\left(  \hat{\vec{q}}_{1}^{\;2}\hat{\vec{q}%
}_{2}^{\;2}\right)  ^{-\frac{1}{2}}\rightarrow\langle A^{\prime}\bar
{A}|\left(  \hat{\vec{q}}_{1}^{\;2}\hat{\vec{q}}_{2}^{\;2}\right)  ^{-\frac
{1}{2}}\hat{\mathcal{O}}~,
\]
\begin{equation}
\;\;\left(  \hat{\vec{q}}_{1}^{\;2}\hat{\vec{q}}_{2}^{\;2}\right)  ^{-\frac
{1}{2}}|\bar{B}^{\prime}B\rangle\rightarrow{\hat{\mathcal{O}}^{-1}}\left(
\hat{\vec{q}}_{1}^{\;2}\hat{\vec{q}}_{2}^{\;2}\right)  ^{-\frac{1}{2}}|\bar
{B}^{\prime}B\rangle\;, \label{kernel transformation}%
\end{equation}
with any non-singular operator $\hat{\mathcal{O}}$. Taking $\hat{\mathcal{O}%
}=\left(  \hat{\vec{q}}_{1}^{\;2}\hat{\vec{q}}_{2}^{\;2}\right)^{1/2}$
we get~(\ref{cal K}) and the right-hand side of the discontinuity~(\ref{discontinuity
representation}) becomes
\begin{equation}
\langle A^{\prime}\bar{A}|\left(  \hat{\vec{q}}_{1}^{\;2}\hat{\vec{q}}%
_{2}^{\;2}\right)  ^{-\frac{1}{2}}e^{Y\hat{{K}}}\left(  \hat{\vec{q}}%
_{1}^{\;2}\hat{\vec{q}}_{2}^{\;2}\right)  ^{-\frac{1}{2}}|\bar{B}^{\prime
}B\rangle\ =\langle A^{\prime}\bar{A}|e^{Y\hat{\mathcal{K}}}\left(  \hat
{\vec{q}}_{1}^{\;2}\hat{\vec{q}}_{2}^{\;2}\right)  ^{-1}|\bar{B}^{\prime
}B\rangle\;. \label{discontinuity in cal K}%
\end{equation}

It is important that, after setting the kernel $\mathcal{K}$ by
Eq.~(\ref{cal K}), which provides conformal invariance of its M\"{o}bius
representation in the LO, an additional transformation with $\hat{\mathcal{O}%
}=1-\alpha_{s}\hat{U}$ is still possible. With NLO accuracy it gives
\[
\hat{\mathcal{K}}\rightarrow\hat{\mathcal{K}}-\alpha_{s}[\hat{\mathcal{K}%
}^{(B)},\hat{U}]~,\;\;\langle A^{\prime}\bar{A}|\rightarrow\langle A^{\prime
}\bar{A}|-\left(  \langle A^{\prime}\bar{A}|\right)  ^{(B)}\alpha_{s}\hat{U},
\]%
\begin{equation}
\left(  \hat{\vec{q}}_{1}^{\;2}\hat{\vec{q}}_{2}^{\;2}\right)  ^{-1}|\bar
{B}^{\prime}B\rangle\rightarrow\left(  \hat{\vec{q}}_{1}^{\;2}\hat{\vec{q}%
}_{2}^{\;2}\right)  ^{-1}|\bar{B}^{\prime}B\rangle+\alpha_{s}\hat{U}\left(
\hat{\vec{q}}_{1}^{\;2}\hat{\vec{q}}_{2}^{\;2}\right)  ^{-1}\left(  |\bar
{B}^{\prime}B\rangle\right)  ^{(B)}. \label{transformation at NLO}%
\end{equation}

It was shown~\cite{Fadin:2009gh} that the
transformation~(\ref{transformation at NLO}) permits one to remove the
discrepancy between the BFKL and the colour dipole kernels (with account of
the correction of the result of Ref.~\cite{Balitsky:2008zz} made in
Ref.~\cite{Balitsky:2009xg}), and to obtain the kernel
\begin{equation}
\hat{\mathcal{K}}^{QC}=\hat{\mathcal{K}}-\alpha_{s}[\hat{\mathcal{K}}%
^{(B)},\hat{U}]~ \label{QC kernel}%
\end{equation}
which is quasi-conformal in the M\"{o}bius representation.

The operator $\hat{U}$ was found as the sum of two pieces, $\hat{U}=\hat{U}%
_{1}+\hat{U}_{2}$. The first piece was found in the momentum space,
\[
\langle\vec{q}_{1},\vec{q}_{2}|\alpha_{s}{\hat{U}}_{1}|\vec{q}_{1}%
^{\,\,\prime},\vec{q}_{2}^{\,\,\prime}\rangle=\frac{\alpha_{s}N_{c}}{2\pi^{2}%
}\Biggl[-\delta(\vec{q}-\vec{q}^{\;\prime})\left(  \frac{\vec{k}}{\vec
{k}^{\,\,2}}-\frac{\vec{q}_{1}}{\vec{q}_{1}^{\,\,2}}\right)  \left(
\frac{\vec{k}}{\vec{k}^{\,\,2}}+\frac{\vec{q}_{2}}{\vec{q}_{2}^{\,\,2}%
}\right)  \ln\vec{k}^{\,\,2}%
\]%
\[
+\delta(\vec{q}_{1}-\vec{q}_{1}^{\;\prime})\delta\left(  \vec{q}_{2}-\vec
{q}_{2}^{\;\prime}\right)  \left(  \int d^{2}l\left(  \frac{1}{\vec{l}%
^{\,\,2}}-\frac{\vec{l}(\vec{l}-\vec{q}_{1})}{2\vec{l}^{\,\,2}(\vec{l}-\vec
{q}_{1})^{2}}-\frac{\vec{l}(\vec{l}-\vec{q}_{2})}{2\vec{l}^{\,\,2}(\vec
{l}-\vec{q}_{2})^{2}}\right)  \ln\vec{l}^{\,\,2}\right.
\]%
\begin{equation}
\left.  -\frac{\pi\beta_{0}}{4N_{c}}\ln\left(  {\vec{q}}_{1}^{\,2}{\vec{q}%
}_{2}^{\,2}\right)  \right)  \Biggr], \label{U1}%
\end{equation}
where $\vec{q}=\vec{q}_{1}+\vec{q}_{2},\;\vec{q}^{\;\prime}=\vec{q}%
_{1}^{\;\prime}+\vec{q}_{2}^{\;\prime},$ $\beta_{0}$ is the first coefficient
of the $\beta$-function, and $\vec{k}=\vec{q}_{1}-\vec{q}_{1}^{\;\prime}$.
Note that the integral over $\vec l$ diverges in $\vec{l}=0$ and, strictly 
speaking, the term with $1/\vec{l}^{\,\,2}$ must be regularized. But in fact 
what we need is the action of the operator $U_1$ on some state,  i.e. the 
integral over $\vec{k}$ of the product of the matrix element and  a 
function of $\vec{k}$, rather than the matrix element itself. 
In this integral  the singularities at $\vec{l}=0$ and $\vec{k}=0$ cancel and 
we get a finite result (evidently, the terms with  $1/\vec{l}^{\,\,2}$ 
and $1/\vec{k}^{\,\,2}$ must be regularized in the same way).
The second part
\[
\langle\vec{r}_{1}\vec{r}_{2}|\alpha_{s}{\hat{U}}_{2M}|\vec{r}_{1}^{\;\prime
}\vec{r}_{2}^{\;\prime}\rangle=\frac{\alpha_{s}N_{c}}{4\pi^{2}}\int d\vec
{r}_{0}\frac{\vec{r}_{12}^{\;2}}{\vec{r}_{01}^{\,\,2}\vec{r}_{02}^{\,\,2}}%
\ln\left(  \frac{\vec{r}_{12}^{\;2}}{\vec{r}_{01}^{\,\,2}\vec{r}_{02}^{\,\,2}%
}\right)
\]%
\begin{equation}
\times\Biggl[\delta(\vec{r}_{11^{\prime}})\delta(\vec{r}_{02^{\prime}}%
)+\delta(\vec{r}_{01^{\prime}})\delta(\vec{r}_{22^{\prime}})-\delta(\vec
{r}_{11^{\prime}})\delta({r}_{22^{\prime}})\Biggr]~, \label{U2}%
\end{equation}
was found in the impact parameter space and in the M\"{o}bius representation,
which is indicated by the subscript $M$ (hereafter $\vec{r}_{ij^{\prime}}%
=\vec{r}_{i}-\vec{r}_{j}^{\;\prime}).$

Thus, the quasi-conformal kernel $\hat{\mathcal{K}}^{QC}$ determined by
Eqs.~(\ref{QC kernel})--(\ref{U2}) was found in the impact parameter space and
in the M\"{o}bius representation. Now its explicit form is known in this space 
and this representation for theories containing fermions and scalars in 
arbitrary representations of the colour group~\cite{Fadin:2010zz}.

Remind that transition to the M\"{o}bius representation means truncation of the
space of states. Therefore, the connection between operators in this
representation and in the complete space of states (i.e. in the ``complete''
representation) is not obvious. In particular, it is not clear if it is
possible to restore ${\hat{U}}_{2}$ (and consequently $\hat{\mathcal{K}}^{QC}%
$) in the complete representation in the momentum space from Eq.~(\ref{U2}). In
the next Section we prove the possibility of such restoration.

\section{Interrelation between complete and M\"{o}\-bi\-us re\-presentations}
\label{sec:interrelation}

The possibility to rebuild the complete kernel from its M\"{o}bius
representation is based on the gauge invariance of the kernel.
Note that this property, together with the gauge invariance of impact factors for
colourless particles, was used for the transition to the M\"{o}bius
representation in Ref.~\cite{Lipatov:1985uk}. Only thanks to this property
the discontinuity~(\ref{discontinuity representation}) can be written using
the M\"{o}bius representation of $\mathcal{\hat{K}}$. Let us remind how the
passage to the M\"{o}bius representation was done.

Gauge invariance of impact factors means that
\begin{equation}
\langle A^{\prime}\bar{A}|\vec{q},0\rangle=\langle A^{\prime}\bar{A}|0,\vec
{q}\rangle=\langle\vec{q},0|\bar{B}^{\prime}B\rangle=\langle0,\vec{q}|\bar
{B}^{\prime}B\rangle=0~, \label{IF gauge invariance}%
\end{equation}
while gauge invariance of the kernel $\hat{K}$ implies the property
\[
{K}_{r}(\vec{q}_{1},\vec{q}_{1}^{\;\prime};\vec{q})|_{{\vec{q}_{1}}=0}={K}%
_{r}(\vec{q}_{1},\vec{q}_{1}^{\;\prime};\vec{q})|_{{\vec{q}_{1}^{\;\prime}}=0}%
\]%
\begin{equation}
={K}_{r}(\vec{q}_{1},\vec{q}_{1}^{\;\prime};\vec{q})|_{{\vec{q}_{1}}=\vec{q}%
}={K}_{r}(\vec{q}_{1},\vec{q}_{1}^{\;\prime};\vec{q})|_{{\vec{q}_{1}%
^{\;\prime}}=\vec{q}}=0\;. \label{K gauge invariance}%
\end{equation}
As we may easily see from Eqs.~(\ref{operator of the BFKL kernel}),~(\ref{real
kernel}), and~(\ref{discontinuity representation}), just these properties
 guarantee  the absence of Coulomb divergences in the discontinuities.

From Eqs.~(\ref{cal K}),~(\ref{real kernel}) and these properties, it also follows
that
\begin{equation}
\langle A^{\prime}\bar{A}|e^{Y\hat{\mathcal{K}}}|\vec{q},0\rangle=\langle
A^{\prime}\bar{A}|e^{Y\hat{\mathcal{K}}}|0,\vec{q}\rangle=0~.
\label{gauge invariance}%
\end{equation}
It means that $\langle A^{\prime}\bar{A}|e^{Y\hat{\mathcal{K}}}|\Psi\rangle=0$
if $\langle\vec{r}_{1},\vec{r}_{2}|\Psi\rangle$ does not depend either on
$\vec{r}_{1}$ or on $\vec{r}_{2}$. Then, Eq.~(\ref{discontinuity in cal K})
shows that one can make the replacement
\[
\langle\vec{r}_{1},\vec{r}_{2}|\bigl(\hat{\vec{q}}_{1}^{\,2}\hat{\vec{q}}%
_{2}^{\,2}\bigr)^{-1}|\bar{B}^{\prime}B\rangle\rightarrow\langle\vec{r}%
_{1},\vec{r}_{2}|\left(  \bigl(\hat{\vec{q}}_{1}^{\,2}\hat{\vec{q}}_{2}%
^{\,2}\bigr)^{-1}|\bar{B}^{\prime}B\rangle\right)  _{M}=\langle\vec{r}%
_{1},\vec{r}_{2}|\bigl(\hat{\vec{q}}_{1}^{\,2}\hat{\vec{q}}_{2}^{\,2}%
\bigr)^{-1}|\bar{B}^{\prime}B\rangle
\]%
\begin{equation}
-\frac{1}{2}\langle\vec{r}_{1},\vec{r}_{1}|\bigl(\hat{\vec{q}}_{1}^{\,2}%
\hat{\vec{q}}_{2}^{\,2}\bigr)^{-1}|\bar{B}^{\prime}B\rangle-\frac{1}{2}%
\langle\vec{r}_{2},\vec{r}_{2}|\bigl(\hat{\vec{q}}_{1}^{\,2}\hat{\vec{q}}%
_{2}^{\,2}\bigr)^{-1}|\bar{B}^{\prime}B\rangle~, \label{BB M}%
\end{equation}
retaining the discontinuity~(\ref{discontinuity representation}). Evidently,
this substitution transfers the state $\bigl(\hat{\vec{q}}_{1}^{\,2}
\hat{\vec{q}}_{2}^{\,2}\bigr)^{-1}|\bar{B}^{\prime}B\rangle$ into the
M\"{o}bius representation. Note that with the requirement of symmetry of
the state with respect to the Reggeon exchange this transformation is unique.
In the momentum space it reads
\[
\langle\vec{q}_{1},\vec{q}_{2}|\bigl(\hat{\vec{q}}_{1}^{\,2}\hat{\vec{q}}%
_{2}^{\,2}\bigr)^{-1}|\bar{B}^{\prime}B\rangle_{M}=\langle\vec{q}_{1},\vec
{q}_{2}|\bigl(\hat{\vec{q}}_{1}^{\,2}\hat{\vec{q}}_{2}^{\,2}\bigr)^{-1}%
|\bar{B}^{\prime}B\rangle
\]%
\begin{equation}
-\frac{1}{2}\biggl(  \delta(\vec{q}_{1}-\vec{q}_{B})\delta(\vec q_{2})+\delta
(\vec{q}_{2}-\vec{q}_{B})\delta(\vec q_{1})\biggr)  \int d\vec{l}_{1}d\vec{l}%
_{2}\langle\vec{l}_{1},\vec{l}_{2}|\bigl(\hat{\vec{q}}_{1}^{\,2}\hat{\vec{q}%
}_{2}^{\,2}\bigr)^{-1}|\bar{B}^{\prime}B\rangle\;. \label{BB M1}%
\end{equation}
Then one can transfer from $\mathcal{K}$ to $\mathcal{K}_{M}$ without
changing the discontinuity. This is done omitting in $\mathcal{K}$ both the
terms which are zero in the M\"{o}bius subspace and the terms whose action on
 any state  in the M\"{o}bius subspace puts  it  out of this
 subspace. Note that the last procedure is not unique. Indeed,
the kernel $\langle\vec{r}_{1},\vec{r}_{2}|\hat
{\mathcal{K}}_M|\vec{r}_{1}^{\;\prime},\vec{r}_{2}^{\;\prime}\rangle$
in the impact parameter space can be written as
\begin{equation}
\langle\vec{r}_{1},\vec{r}_{2}|\hat{\mathcal{K}}_{M}|\vec{r}_{1}^{\;\prime
},\vec{r}_{2}^{\;\prime}\rangle=\langle\vec{r}_{1},\vec{r}_{2}|\hat
{\mathcal{K}}|\vec{r}_{1}^{\;\prime},\vec{r}_{2}^{\;\prime}\rangle_{t}%
-f_1(\vec{r}_{11^{\prime}},  \vec{r}_{12^{\prime}})-f_2(\vec{r}_{21^{\prime}},  \vec{r}_{22^{\prime}})~,
\label{K M impact}%
\end{equation}
where the subscript $t$ means omitting the terms proportional to
$\delta(\vec{r}_{1^{\prime}2^{\prime}})$, and  functions $f_1$ and
$f_2$ are restricted (besides the absence of the terms proportional to
$\delta(\vec{r}_{1^{\prime}2^{\prime}})$) by the requirement
\begin{equation}
f_1(\vec{r}_{01^{\prime}},  \vec{r}_{02^{\prime}})+f_2(\vec{r}_{01^{\prime}},  \vec{r}_{02^{\prime}})=\langle\vec{r}_0,\vec{r}_0|\hat{\mathcal{K}}|\vec{r}_{1}^{\;\prime}%
,\vec{r}_{2}^{\;\prime}\rangle_{t}\, .
\label{K M impact requirement}%
\end{equation}
But the  uncertainty in the choice of $f_1$ and $f_2$  plays no role
due to the symmetry of the impact factors $\langle A^{\prime
}\bar{A}|$ with respect to the Reggeon exchange. Indeed, if we have
two sets of functions $f^{(1)}_i$ and $f^{(2)}_i, \, i=1,2$,
satisfying (\ref{K M impact requirement}), then the difference
$$\left[f^{(1)}_1(\vec{r}_{11^{\prime}},  \vec{r}_{12^{\prime}})+f^{(1)}_2(\vec{r}_{21^{\prime}},  \vec{r}_{22^{\prime}})\right]- \left[f^{(2)}_1(\vec{r}_{11^{\prime}},  \vec{r}_{12^{\prime}})+f^{(2)}_2(\vec{r}_{21^{\prime}},  \vec{r}_{22^{\prime}})\right]$$ is antisymmetric with respect to replacement
$\vec{r}_1\leftrightarrow \vec{r}_2 $. In fact, this uncertainty can
be used for the simplification of
$\langle\vec{r}_{1},\vec{r}_{2}|\hat{\mathcal{K}}_{M}|\vec{r}_{1}^{\;\prime
},\vec{r}_{2}^{\;\prime}\rangle$. On the other hand, if one does not
like the uncertainty,  it  can be removed imposing the requirement of
the corresponding symmetry on the kernel, so that in the following
we will not pay attention to it.

Thus, in the impact parameter space the M\"{o}bius representation of the kernel
$\mathcal{K}$ is unambiguously constructed from the complete one, assuming the
symmetry with respect to the Reggeon exchange (which follows from the boson
nature of Reggeons). Evidently, this statement is valid for any operator
defined both in the momentum and the coordinate spaces.

Note that, strictly speaking, operators in the M\"{o}bius
representation are not defined in the momentum space. The reason is
that in the impact parameter space they can be singular at
$\vec{r}_{1^{\prime}2^{\prime}}=0$, so that their direct
transformation into the momentum space can be impossible.  Using
translation invariance, we can formally write
\[
\langle\vec{q}_{1},\vec{q}_{2}|\hat{\mathcal{K}}_{M}|\vec{q}_{1}^{\;\prime
},\vec{q}_{2}^{\;\prime}\rangle\!=\!\int\frac{d\vec{r}_{1}}{2\pi}\frac{d\vec
{r}_{2}}{2\pi}\frac{d\vec{r}_{1}^{\;\prime}}{2\pi}\frac{d\vec{r}_{2}%
^{\;\prime}}{2\pi}e^{-i\vec{q}_{1}\vec{r}_{1}-i\vec{q}_{2}\vec{r}_{2}+i\vec
{q}_{1}^{\;\prime}\vec{r}_{1}^{\;\prime}+i\vec{q}_{2}^{\;\prime}\vec{r}%
_{2}^{\;\prime}}\langle\vec{r}_{1},\vec{r}_{2}|\hat{\mathcal{K}}_{M}|\vec
{r}_{1}^{\;\prime},\vec{r}_{2}^{\;\prime}\rangle
\]%
\begin{equation}
=\delta(\vec{q}_{1}+\vec{q}_{2}-\vec{q}_{1}^{\;\prime}-\vec{q}_{2}^{\;\prime
})\mathcal{K}_{M}(\vec{q}_{1},\vec{q}_{2};\vec{k})\;,
\label{KM formal in q-space}%
\end{equation}
where $\vec{k}=\vec{q}_{1}-\vec{q}_{1}^{\;\prime}=\vec{q}_{2}^{\;\prime}%
-\vec{q}_{2}$ and
\begin{equation}
\mathcal{K}_{M}(\vec{q}_{1},\vec{q}_{2};\vec{k})=\int\frac{d\vec
{r}_{11^{\prime}}}{2\pi}\frac{d\vec{r}_{22^{\prime}}}{2\pi}{d\vec
{r}_{1^{\prime}2^{\prime}}}e^{-i\vec{q}_{1}\vec{r}_{11^{\prime}}-i\vec{q}%
_{2}\vec{r}_{22^{\prime}}-i\vec{k}\vec{r}_{1^{\prime}2^{\prime}}}\langle
\vec{r}_{1},\vec{r}_{2}|\hat{\mathcal{K}}_{M}|\vec{r}_{1}^{\;\prime},\vec
{r}_{2}^{\;\prime}\rangle \;,\label{KM direct in q-space}%
\end{equation}
(not to be confused with
${K}_{r}(\vec{q}_{1},\vec{q}_{1}^{\;\prime};\vec {q})$). But the integral over $\vec{r}_{1^{\prime}2^{\prime}}$ in ~(\ref{KM direct in
q-space}) can be divergent because of  singularities of the type $\ln^n
\vec{r}_{1^{\prime}2^{\prime}}^{\,2}/\vec{r}_{1^{\prime}2^{\prime}}^{\,2}$.  It is useful to understand that
singularities at $\vec{r}_{1^{\prime}2^{\prime}}=0$ are related with the
growth of operators in momentum space at large $\vec k^{\:2}$. Thus,
the NLO BFKL kernel in momentum space contains terms with $\ln \vec
k^{\:2}$ and  $\ln^2 \vec k^{\:2}$ behaviour  at large $\vec k^{\:2}$.  At fixed $\vec{r}_{1^{\prime}2^{\prime}}\neq0$ we have
\[
\int\frac{d\vec{k}}{2\pi}e^{i\vec{k}\,\vec{r}_{1^{\prime}2^{\prime}}}\ln({\vec{k}^{\,2}})=-\frac
{2}{\vec{r}_{1^{\prime}2^{\prime}}^{\,2}}\, ,
\]
\begin{equation}
\int\frac{d\vec{k}}{2\pi}e^{i\vec{k}\,\vec{r}_{1^{\prime}2^{\prime}}}\ln^2({\vec{k}^{\,2}})
=\frac{4}{\vec{r}_{1^{\prime}2^{\prime}}^{\,2}}\left(\ln\left(
\frac{\vec{r}_{1^{\prime}2^{\prime}}^{\;2}}{4}\right)-2\psi(1)\right)\;\;\; . \label{int ln k^2}%
\end{equation}
This result can be obtained, for example, writing
$$\ln^n\vec{k}^{\;2}
=(-1)^n\frac{d^n}{d\alpha^n}\left.(\vec{k}^{\;2})^{-\alpha}\right|_{\alpha=0}
$$ 
and using the equality
\begin{equation}
\int\frac{d\vec{k}}{2\pi}(\vec{k}^{\;2})^{-\alpha}e^{i\vec{k}\,\vec{r}}%
=\frac{2}{\vec{r}^{\,2}}\frac{\Gamma(1-\alpha)}{\Gamma(1+\alpha)}\alpha\left(
\frac{\vec{r}^{\;2}}{4}\right)  ^{\alpha}. \label{int ks -alpha}%
\end{equation}
In fact,   the limits $\alpha\rightarrow0$ and
$\vec{r}^{\;2}\rightarrow0$ are not interchangeable. It means that
result~(\ref{int ln k^2}) cannot be used at arbitrary small
$\vec{r}_{1^{\prime}2^{\prime}}^{\;2}$.  In the M\"{o}bius
representation small $\vec{r}_{1^{\prime}2^{\prime}}^{\;2}$ are
unimportant and the result  (\ref{int ln k^2}) is used everywhere.
But in the integral (\ref{KM direct in q-space}) the singularity at
$\vec{r}_{1^{\prime}2^{\prime}}^{\;2}$ must be regularized. From the
consideration above it is natural to use in~(\ref{KM direct in
q-space}), instead of $(1/\vec{r}^{\;2})$ and $(\ln
\vec{r}^{\;2}/\vec{r}^{\;2})$, the regularized functions
$(1/\vec{r}^{\;2})_{R}$ and $(\ln \vec{r}^{\;2}/\vec{r}^{\;2})_R$
which  make possible the inverse Fourier transform,
\[
\int\frac{d\vec{r}}{2\pi}\left(\frac{1}{\vec{r}^{\;2}}\right)_{R} e^{-i\vec{k}\,\vec{r}%
}=-\frac{1}{2}\ln({\vec{k}^{\,2}})\;,  %
\]
\begin{equation}
\int\frac{d\vec{r}}{2\pi}\left(\frac{1}{\vec{r}^{\;2}}\left[\ln\left(
\frac{\vec{r}^{\:2}}{4}\right)-2\psi(1)\right]\right)_{R}e^{-i\vec{k}\,\vec{r}%
}=\frac{1}{4}\ln^2({\vec{k}^{\,2}})\;.  \label{int xs r}%
\end{equation}
Since
\[
\int\frac{d\vec{r}}{2\pi}\frac{e^{-i\vec{k}\,\vec{r}}}{\vec{r}^{\;2}}%
\theta(\vec{r}^{\;2}-c^{2})|_{c\rightarrow0}=-\frac{1}{2}\left(\ln\left(\frac{\vec{k}^{\,2}}{4}\right)-2\psi(1)+\ln
c^{2}\right)\, ,
\]
\begin{equation}
\int\frac{d\vec{r}}{2\pi}\frac{e^{-i\vec{k}\,\vec{r}}}{\vec{r}^{\;2}}\ln\vec{r}^{\;2}%
\theta(\vec{r}^{\;2}-c^{2})|_{c\rightarrow
0}=\frac{1}{4}\left(\left(\ln\left(\frac{\vec{k}^{\,2}}{4}\right)-2\psi(1)\right)^2-\ln^2
c^{2}\right)\, ,
\end{equation}
this can be done defining $1/(\vec{r}^{\;2})_{R}$  and  $(\ln
\vec{r}^{\;2}/\vec{r}^{\;2})$ at
$\vec{r}^{\;2}\rightarrow0$ in the following way:
\[
\int\frac{d\vec{r}}{2\pi}\left(\frac{1}{\vec{r}^{\;2}}\right)_{R}\theta(c^{2}-\vec
{r}^{\;2})|_{c\rightarrow0}=\frac{1}{2}\left(\ln c^{2}-2\psi(1)-\ln4\right)\;,
\]
\begin{equation}
\int\frac{d\vec{r}}{2\pi}\left(\frac{\ln\vec{r}^{\;2}}{\vec{r}^{\;2}}\right)_{R}\theta(c^{2}-\vec
{r}^{\;2})|_{c\rightarrow0}=\frac{1}{4}\left(\ln^2 c^{2}-(2\psi(1)+\ln4)^2\right)\;.
\label{definition xs r}%
\end{equation}
For definiteness, let us accept that (\ref{KM direct in q-space}) is defined 
with such regularization. But, in fact, the choice of a regularization is not 
important for the restoration of the complete kernel from the M\"{o}bius one. 
Indeed, since any regularization concerns only the region  
$\vec{r}_{1^{\prime}2^{\prime}}\rightarrow 0$, any change of regularization 
has an influence only on terms not depending on $\vec k$. Therefore, denoting
\begin{equation}
\langle\vec{q}_{1},\vec{q}_{2}|\hat{\mathcal{K}}|\vec{q}_{1}^{\;\prime}%
,\vec{q}_{2}^{\;\prime}\rangle=\delta(\vec{q}_{1}+\vec{q}_{2}-\vec{q}%
_{1}^{\;\prime}-\vec{q}_{2}^{\;\prime})\mathcal{K}(\vec{q}_{1},\vec{q}%
_{2};\vec{k})\;, \label{kernel cal K}%
\end{equation}
we obtain from Eq.~(\ref{K M impact})
\[
\mathcal{K}(\vec{q}_{1},\vec{q}_{2};\vec{k})_{-}=\mathcal{K}_{M}(\vec{q}%
_{1},\vec{q}_{2};\vec{k})_{-} +\delta(\vec{q_{2}})\int{d\vec
{r}_{11^{\prime}}}{d\vec{r}_{12^{\prime}}}e^{-i(\vec{q}_{1}-\vec{k})\vec{r}_{11^{\prime}
}-i\vec{k}\vec{r}_{12^{\prime}}}f_1(\vec{r}_{11^{\prime}},\vec{r}_{12^{\prime}})
\]%
\begin{equation}
+\delta(\vec{q_{1}})\int{d\vec
{r}_{21^{\prime}}}{d\vec{r}_{22^{\prime}}}e^{-i(\vec{q}_{2}+\vec{k})\vec{r}_{22^{\prime}
}+i\vec{k}\vec{r}_{21^{\prime}}}f_2(\vec{r}_{21^{\prime}},\vec{r}_{22^{\prime}})~,
\label{K M momentum}%
\end{equation}
where the subscript $\mathcal{K}(\vec{q}_{1},\vec{q}_{2};\vec{k})_{-}$ means
$\mathcal{K}(\vec{q}_{1},\vec{q}_{2};\vec{k})$ without terms independent of
the third argument in $\mathcal{K}(\vec{q}_{1},\vec{q}_{2};\vec{k})$.

As was stated before, the M\"{o}bius representation of the BFKL kernel is
unambiguously constructed from the complete one, by requiring the symmetry with
respect to the Reggeon exchange. But the inverse statement is also valid.
The M\"{o}bius representation of the BFKL kernel totally defines the complete
kernel symmetric with respect to Reggeon exchange. For the validity of this
statement two properties of the complete BFKL kernel are important. The
first one is its gauge invariance which gives (see Eqs.~(\ref{cal K}),~(\ref{K
gauge invariance}) and~(\ref{kernel cal K}))
\begin{equation}
\mathcal{K}_r(\vec{q}_{1},\vec{q}_{2};\vec{q}_{1})=\mathcal{K}_r(\vec{q}_{1}%
,\vec{q}_{2};-\vec{q}_{2})=0\;. \label{gauge invariance for cal  K}%
\end{equation}
And the second one is the absence
of terms proportional to $\delta(\vec{q_{1}})$ or $\delta(\vec{q_{2}})$ in
the kernel. This fixes the residual freedom connected with such terms.

These  properties provide (with account of the symmetrization discussed
above) the uniqueness of the restoration of the total kernel from its
M\"{o}bius representation. Indeed, if there were two different complete kernels
$\mathcal{K}^{(1)}$ and $\mathcal{K}^{(2)}$ with the same M\"{o}bius
representation, then the M\"{o}bius representation for their difference would
be zero. It follows from Eq.~(\ref{K M momentum}) that in this case it must be
\begin{equation}
\mathcal{K}^{(1)}(\vec{q}_{1},\vec{q}_{2};\vec{k})_{-}-\mathcal{K}^{(2)}%
(\vec{q}_{1},\vec{q}_{2};\vec{k})_{-}=0, \label{K 1 - - K 2 - =0}%
\end{equation}
i.e. they can differ only in terms independent of $\vec{k}$. On the other
hand, gauge invariance requires turning the difference
 $\mathcal{K}^{(1)}%
(\vec{q}_{1},\vec{q}_{2};\vec{k})_{-}-\mathcal{K}^{(2)}(\vec{q}_{1},\vec
{q}_{2};\vec{k})_{-}$ into zero at $\vec{k}=\vec{q}_{1}$ and at
$\vec{k}=-\vec{q}_{2}$. Therefore, it is zero identically.

Thus, the uniqueness of the restoration of $\hat{\mathcal{K}}$ from 
$\hat{\mathcal{K}}_{M}$ is proven. This proof and~(\ref{K M momentum})  
give the way to perform the restoration.

\section{Restoration of the operator $U_{2}$ from its M\"{o}\-bius form}
\label{sec:restoration}

Let us demonstrate the restoration of the complete
operator from its M\"{o}bius form on the example of the operator
$U_{2}$ given in Eq.~(\ref{U2}). First of all, it is necessary to note that
$\hat{U}$ in the transformation~(\ref{transformation at NLO}) cannot be
arbitrary if we want to conserve the possibility to use the M\"{o}bius
representation after this transformation. Indeed, in this case the
transformation~(\ref{transformation at NLO}) must conserve the gauge
invariance of the impact factor
$\langle A^{\prime}\bar{A}|$ and kernel $\hat{\mathcal{K}}$. Therefore,
$\hat{U}$ must be gauge invariant in the same way as $\hat{\mathcal{K}}$.
Moreover, without any loss of generality we can consider that it has no terms
proportional to $\delta(\vec{q_{1}})$ or $\delta(\vec{q_{2}})$ in the momentum
space, since such terms do not contribute to the
discontinuity~(\ref{discontinuity representation}). In other words,
$\hat{U}$ has the same properties as $\hat{\mathcal{K}}$.
For the part $\hat{U}_{1}$ these properties are easily seen from Eq.~(\ref{U1}).
It means that $\hat{U}_{2}$ has the same properties and therefore can be
unambiguously restored from Eq.~(\ref{U2}).

To do that, let us first transfer Eq.~(\ref{U2}) into the momentum space. As was
pointed above, generally an operator $\mathcal{\hat{O}}$ in the M\"{o}bius
representation in the impact parameter space can contain non-integrable singularities  at $\vec {r}_{1^{\prime}2^{\prime}}=0$ which require regularization. But, as we may notice from Eq.~(\ref{U2}), $\langle\vec{r}%
_{1}\vec{r}_{2}|{\hat{U}}_{2M}|\vec{r}_{1}^{\;\prime}\vec{r}_{2}^{\;\prime
}\rangle$ does not have explicit singularities at $\vec{r}_{1^{\prime
}2^{\prime}}=0$, though its separate parts are divergent. Therefore we will
treat the sum of these parts and calculate directly
$\langle\vec{q}_{1}\vec{q}_{2}|{\hat{U}}_{2M}|\vec{q}_{1}^{\;\prime}\vec
{q}_{2}^{\;\prime}\rangle$.

Here we meet technical problems related to the separation of real and virtual
parts, as usually occurs when one operates with the BFKL kernel. Again looking at
Eq.~(\ref{U2}),  we observe that there are terms in $\langle\vec{r}_{1}\vec{r}_{2}|
{\hat{U}}_{2M}|\vec{r}_{1}^{\;\prime}\vec{r}_{2}^{\;\prime}\rangle$ which
have ultraviolet singularities at $\vec{r}_{01}=0$ and $\vec{r}_{02}=0$ that
cancel in their sum. We will treat them together.
But this is not the only problem. Another problem is that,
quite analogously to $\mathcal{K}(\vec{q}_{1},\vec{q}_{2};\vec{k})$, ${U}%
_{2}(\vec{q}_{1},\vec{q}_{2};\vec{k})$ (defined by Eq.~(\ref{kernel cal K}) with
the substitution ${\mathcal{K}}\rightarrow U_{2}$) contains singularities at
$\vec{k}=0$ in the virtual (proportional to $\delta(\vec{k})$) and the real
parts. They cancel each other; but to make this cancellation evident one needs
to write the coefficient of $\delta(\vec{k})$ in an integral form.

Defining ${U}_{2M}(\vec{q}_{1},\vec{q}_{2};\vec{k})$ according
to Eqs.~(\ref{KM formal in q-space}) and~(\ref{KM direct in q-space}) with the
substitution $\hat{\mathcal{K}}_{M}$$\rightarrow\hat{U}_{2M}$,
using Eq.~(\ref{U2}), and integrating the delta-functions, we obtain
\[
\alpha_{s}{U}_{2M}(\vec{q}_{1},\vec{q}_{2};\vec{k})=\frac{\alpha_{s}N_{c}%
}{4\pi^{2}}\int\frac{d\vec{r}_{1}}{2\pi}\frac{d\vec{r}_{2}}{2\pi}\frac{\vec
{r}_{12}^{\;2}}{\vec{r}_{1}^{\,\,2}\vec{r}_{2}^{\,\,2}}\ln\left(  \frac
{\vec{r}_{12}^{\;2}}{\vec{r}_{1}^{\,\,2}\vec{r}_{2}^{\,\,2}}\right)
\]%
\begin{equation}
\times\left[  e^{-i\vec{k}\,\vec{r}_{1}-i\vec{q}_{2}\,\vec{r}_{2}}%
+e^{-i\vec{q}_{1}\,\vec{r}_{1}+i\vec{k}\,\vec{r}_{2}}-e^{-i\vec{k}\,\vec
{r}_{12}}\right] \;. \label{U2M direct in q-space}%
\end{equation}
Let us divide $\hat{U}_{2}$ into two pieces, $\hat{U}_{2}=\hat{U}_{2}^{r}%
+\hat{U}_{2}^{v}$; making in Eq.~(\ref{U2M direct in q-space}) the following
decomposition:
\[
\frac{\vec{r}_{12}^{\;2}}{\vec{r}_{1}^{\,\,2}\vec{r}_{2}^{\,\,2}}\ln\left(
\frac{\vec{r}_{12}^{\;2}}{\vec{r}_{1}^{\,\,2}\vec{r}_{2}^{\,\,2}}\right)
=\frac{1}{\vec{r}_{1}^{\,\,2}}\ln\left(  \frac{\vec{r}_{12}^{\;2}}{\vec{r}%
_{2}^{\,\,2}}\right)  +\frac{1}{\vec{r}_{2}^{\,\,2}}\ln\left(  \frac{\vec
{r}_{12}^{\;2}}{\vec{r}_{1}^{\,\,2}}\right)  -2\frac{\vec{r}_{1}\vec{r}_{2}%
}{\vec{r}_{1}^{\,\,2}\vec{r}_{2}^{\,\,2}}\ln\left(  \frac{\vec{r}_{12}^{\;2}%
}{\vec{r}_{1}^{\,\,2}\vec{r}_{2}^{\,\,2}}\right)
\]%
\begin{equation}
+\frac{1}{\vec{r}_{1}^{\,\,2}}\ln\left(  \frac{1}{\vec{r}_{1}^{\,\,2}}\right)
+\frac{1}{\vec{r}_{2}^{\,\,2}}\ln\left(  \frac{1}{\vec{r}_{2}^{\,\,2}}\right)\;,
\label{decomposition}%
\end{equation}
the first three terms in the decomposition correspond to $\hat{U}%
_{2}^{r}$ and the last two ones correspond to $\hat{U}_{2}^{v}$. Then
${U}_{2M}^{r}(\vec{q}_{1},\vec{q}_{2};\vec{k})$ is calculated using the
integrals
\[
\int\frac{d\vec{r}_{1}}{2\pi}\frac{d\vec{r}_{2}}{2\pi}\frac{1}{\vec{r}%
_{1}^{\;2}}\ln\left(  \frac{\vec{r}_{12}^{\;2}}{\vec{r}_{2}^{\;2}}\right)
e^{-i\vec{a}\,\vec{r}_{1}-i\vec{b}\,\vec{r}_{2}}=\frac{1}{\vec{b}^{\,2}}%
\ln\left(  \frac{(\vec{a}+\vec{b})^{2}}{\vec{a}^{\,2}}\right)  ,\;\;
\]
\[
\int\frac{d\vec{r}_{1}}{2\pi}\frac{d\vec{r}_{2}}{2\pi}\frac{\vec{r}_{1}\vec
{r}_{2}}{\vec{r}_{1}^{\;2}\vec{r}_{2}^{\;2}}\ln\left(  \frac{\vec{r}%
_{12}^{\;2}}{\vec{r}_{1}^{\;2}}\right)  e^{-i\vec{a}\,\vec{r}_{1}-i\vec
{b}\,\vec{r}_{2}}=-\,\frac{\vec{a}\vec{b}}{\vec{a}^{\,2}\vec{b}^{\,2}}%
\ln\left(  \frac{(\vec{a}+\vec{b})^{2}}{\vec{b}^{\,2}}\right)  ,\;\;
\]
\begin{equation}
\int\frac{d\vec{r}}{2\pi}e^{-i\vec{a}\,\vec{r}}\frac{\vec{r}}{\vec{r}^{\,2}%
}=\frac{-i\vec{a}}{\vec{a}^{\,2}}\;,\label{int k/k^2 ln k/k^2}
\end{equation}
\[
\int\frac{d\vec{r}}{2\pi}e^{-i\vec
{a}\,\vec{r}}\frac{\vec{r}}{\vec{r}^{\,2}}\ln({\vec{r}^{\,2}})=\frac{-i\vec
{a}}{\vec{a}^{\,2}}\left(  2\psi(1)-\ln\left(  \frac{\vec{a}^{\,2}}{4}\right)
\right)\;,  %
\]
with the result
\[
\alpha_{s}{{U}}_{2M}^{r}(\vec{q}_{1},\vec{q}_{2},\vec{k})=\frac{\alpha
_{s}N_{c}}{4\pi^{2}}\left[  \frac{2}{\vec{k}^{\,\,2}}\ln(\vec{k}%
^{\,\,2})+\frac{1}{\vec{q}_{1}^{\,\,2}}\ln\left(  \frac{\vec{q}_{1}%
^{\;\prime\,2}}{\vec{k}^{\;2}}\right)  \right.
\]%
\[
\left.  +\frac{1}{\vec{q}_{2}^{\,\,2}}\ln\left(  \frac{\vec{q}_{2}%
^{\;\prime\,2}}{\vec{k}^{\;2}}\right)  +\frac{1}{\vec{k}^{\,\,2}}\ln\left(
\frac{\vec{q}_{1}^{\;\prime\,2}\vec{q}_{2}^{\;\prime\,2}}{\vec{q}_{1}%
^{\;2}\vec{q}_{2}^{\;2}}\right)  -2\frac{\vec{q}_{1}\vec{k}}{\vec{k}%
^{\,\,2}\vec{q}_{1}^{\,\,2}}\ln\left(  \vec{q}_{1}^{\;\prime\,2}\right)
+2\frac{\vec{q}_{2}\vec{k}}{\vec{k}^{\,\,2}\vec{q}_{2}^{\,\,2}}\ln\left(
\vec{q}_{2}^{\;\prime\,2}\right)  \right.
\]%
\begin{equation}
\left.  -2\left(  \psi(1)+\ln2\right)  \left(  \frac{2}{\vec{k}^{\,\,2}}%
-\frac{2\vec{q}_{1}\vec{k}}{\vec{q}_{1}^{\;2}\vec{k}^{\,\,2}}+\frac{2\vec
{q}_{2}\vec{k}}{\vec{q}_{2}^{\;2}\vec{k}^{\,\,2}}\right)  \right]  ~,
\label{U2 r momentum}%
\end{equation}
where $\vec{q}_{1}^{\;\prime}=\vec{q}_{1}-\vec k,\;\vec{q}_{2}^{\;\prime
}=\vec{q}_{2}+\vec{k}$.

The last two terms into the decomposition~(\ref{decomposition})
after integration~(\ref{U2M direct in q-space}) give contributions
proportional to $\delta (\vec{q}_{1})$, $\delta(\vec{q}_{2})$ and
$\delta(k)$ with divergent coefficients. The terms proportional to
$\delta(\vec{q}_{1})$ and $\delta (\vec{q}_{2})$ can be omitted. The
coefficient of $\delta(k)$ is presented in an integral form using
the trick~\cite{Fadin:2006ha}
\[
\int{d\vec{r}}\,\frac{1}{\vec{r}^{\;2}}\ln\left(  \frac{1}{\vec{r}^{\;2}%
}\right)  \left(  e^{i\vec{q}_{1}\,\vec{r}}+e^{i\vec{q}_{2}\,\vec{r}%
}-2\right)
\]%
\[
=\int d\vec{l}\,\frac{d\vec{r}_{1}}{2\pi}\,\frac{d\vec{r}_{2}}{2\pi}%
\,\frac{\vec{r}_{1}\vec{r}_{2}}{\vec{r}_{1}^{\;2}\vec{r}_{2}^{\;2}}%
e^{-i\,\vec{l}\,(\vec{r}_{1}+\vec{r}_{2})}\Bigl[e^{i\vec{q}_{1}\,\vec{r}_{1}%
}\ln({\vec{r}_{2}^{\;2}})+e^{i\vec{q}_{2}\,\vec{r}_{2}}\ln({\vec{r}_{1}^{\;2}%
})-\ln({\vec{r}_{1}^{\;2}}{\vec{r}_{2}^{\;2}})\Bigr]
\]%
\begin{equation}
=\int d\vec{l}\,\Bigl[2\psi(1)+\ln4-\ln({\vec{l}^{\;2}})\Bigr]\left(  \frac
{2}{\vec{l}^{\,\,2}}-\,\frac{\vec{l}(\vec{l}-\vec{q}_{1})}{\vec{l}%
^{\,\,2}(\vec{l}-\vec{q}_{1})^{2}}-\frac{\vec{l}(\vec{l}-\vec{q}_{2})}{\vec
{l}^{\,\,2}(\vec{l}-\vec{q}_{2})^{2}}\right) \;.\;\;
\end{equation}
At last, the real part~(\ref{U2 r momentum}) must be changed by adding terms
independent of $\vec{k}$ in such a way that it becomes zero at $\vec{q}%
_{1}^{\,\,\prime}=0$ and $\vec{q}_{2}^{\,\,\prime}=0$. The result is
\[
\langle\vec{q}_{1},\vec{q}_{2}|\alpha_{s}{\hat{U}}_{2}|\vec{q}_{1}%
^{\,\,\prime},\vec{q}_{2}^{\,\,\prime}\rangle=\delta(\vec{q}_{11^{\prime}%
}+\vec{q}_{22^{\prime}})\frac{\alpha_{s}N_{c}}{4\pi^{2}}\left[  \frac{2}%
{\vec{k}^{\,\,2}}\ln(\vec{k}^{\,\,2})+\frac{1}{\vec{q}_{1}^{\,\,2}}\ln\left(
\frac{\vec{q}_{1}^{\;\prime\,2}\vec{q}_{2}^{\;2}}{\vec{k}^{\;2}\vec{q}^{\;2}%
}\right)  \right.
\]%
\[
\left.  +\frac{1}{\vec{q}_{2}^{\,\,2}}\ln\left(  \frac{\vec{q}_{2}%
^{\;\prime\,2}\vec{q}_{1}^{\;2}}{\vec{k}^{\;2}\vec{q}^{\;2}}\right)  +\frac
{1}{\vec{k}^{\,\,2}}\ln\left(  \frac{\vec{q}_{1}^{\;\prime\,2}\vec{q}%
_{2}^{\;\prime\,2}}{\vec{q}_{1}^{\;2}\vec{q}_{2}^{\;2}}\right)  -2\frac
{\vec{q}_{1}\vec{k}}{\vec{k}^{\,\,2}\vec{q}_{1}^{\,\,2}}\ln\left(  \vec{q}%
_{1}^{\;\prime\,2}\right)  +2\frac{\vec{q}_{2}\vec{k}}{\vec{k}^{\,\,2}\vec
{q}_{2}^{\,\,2}}\ln\left(  \vec{q}_{2}^{\;\prime\,2}\right)  \right.
\]%
\[
\left.  -2\frac{\vec{q}_{1}\vec{q}_{2}}{\vec{q}_{1}^{\,\,2}\vec{q}_{2}%
^{\,\,2}}\ln(\vec{q}^{\;2})\right]  -\frac{\alpha_{s}N_{c}}{4\pi^{2}}%
\,\delta(\vec{q}_{22^{\prime}})\delta\left(  \vec{q}_{11^{\prime}}\right)
\int d\vec{l}\ln\vec{l}^{\,\,2}\left(  \frac{2}{\vec{l}^{\,\,2}}-\frac{\vec
{l}(\vec{l}-\vec{q}_{1})}{\vec{l}^{\,\,2}(\vec{l}-\vec{q}_{1})^{2}}\right.
\]%
\begin{equation}
\left.  -\frac{\vec{l}(\vec{l}-\vec{q}_{2})}{\vec{l}^{\,\,2}(\vec{l}-\vec
{q}_{2})^{2}}\right)  -\left(  \psi(1)+\ln2\right)  \langle\vec{q}_{1},\vec
{q}_{2}|{\hat{\mathcal{K}}}^{B}|\vec{q}_{1}^{\,\,\prime},\vec{q}%
_{2}^{\,\,\prime}\rangle~, \label{U2 momentum}%
\end{equation}
where
\[
\langle\vec{q}_{1},\vec{q}_{2}|{\hat{\mathcal{K}}}^{(B)}|\vec{q}%
_{1}^{\,\,\prime},\vec{q}_{2}^{\,\,\prime}\rangle=\delta(\vec{q}_{11^{\prime}%
}+\vec{q}_{22^{\prime}})\frac{\alpha_{s}N_{c}}{2\pi^{2}}\left[  \frac{2}%
{\vec{k}^{\,\,2}}-2\frac{\vec{q}_{1}\vec{k}}{\vec{k}^{\,\,2}\vec{q}%
_{1}^{\,\,2}}+2\frac{\vec{q}_{2}\vec{k}}{\vec{k}^{\,\,2}\vec{q}_{2}^{\,\,2}%
}\right.
\]%
\begin{equation}
\left.  -2\frac{\vec{q}_{1}\vec{q}_{2}}{\vec{q}_{1}^{\,\,2}\vec{q}_{2}%
^{\,\,2}}-\delta(\vec{k})\int d\vec{l}\left(  \frac{2}{\vec{l}^{\,\,2}}%
-\frac{\vec{l}(\vec{l}-\vec{q}_{1})}{\vec{l}^{\,\,2}(\vec{l}-\vec{q}_{1})^{2}%
}-\frac{\vec{l}(\vec{l}-\vec{q}_{2})}{\vec{l}^{\,\,2}(\vec{l}-\vec{q}%
_{2})^{\,2}}\right)  \right]  . \label{K B  momentum}%
\end{equation}
Evidently, the last term in Eq.~(\ref{U2 momentum}) does not contribute to the
commutator $[{\hat{\mathcal{K}}}^{(B)},{\hat{U}}_{2}]$ and therefore can be omitted.

For the full operator $\hat{U}=\hat{U}_{1}+\hat{U}_{2}$ we have
from Eqs.~(\ref{U1}) and~(\ref{U2 momentum})
\[
\langle\vec{q}_{1},\vec{q}_{2}|\alpha_{s}{\hat{U}}|\vec{q}_{1}^{\,\,\prime
},\vec{q}_{2}^{\,\,\prime}\rangle=\delta(\vec{q}_{11^{\prime}}+\vec
{q}_{22^{\prime}})\frac{\alpha_{s}N_{c}}{4\pi^{2}}\left[  \frac{1}{\vec{q}%
_{1}^{\,\,2}}\ln\left(  \frac{\vec{q}_{1}^{\;\prime\,2}\vec{q}_{2}^{\;2}}%
{\vec{k}^{\;2}\vec{q}^{\;2}}\right)  +\frac{1}{\vec{q}_{2}^{\,\,2}}\ln\left(
\frac{\vec{q}_{2}^{\;\prime\,2}\vec{q}_{1}^{\;2}}{\vec{k}^{\;2}\vec{q}^{\;2}%
}\right)  \right.
\]%
\[
\left.  +\frac{1}{\vec{k}^{\,\,2}}\ln\left(  \frac{\vec{q}_{1}^{\;\prime
\,2}\vec{q}_{2}^{\;\prime\,2}}{\vec{q}_{1}^{\;2}\vec{q}_{2}^{\;2}}\right)
-\frac{2\vec{q}_{1}\vec{k}}{\vec{k}^{\,\,2}\vec{q}_{1}^{\,\,2}}\ln\left(
\frac{\vec{q}_{1}^{\;\prime\,2}}{\vec{k}^{\,\,2}}\right)  +\frac{2\vec{q}%
_{2}\vec{k}}{\vec{k}^{\,\,2}\vec{q}_{2}^{\,\,2}}\ln\left(  \frac{\vec{q}%
_{2}^{\;\prime\,2}}{\vec{k}^{\,\,2}}\right)  -\frac{2\vec{q}_{1}\vec{q}_{2}%
}{\vec{q}_{1}^{\,\,2}\vec{q}_{2}^{\,\,2}}\ln\left(  \frac{\vec{q}^{\;2}}%
{\vec{k}^{\,\,2}}\right)  \right]
\]%
\begin{equation}
-\frac{\alpha_{s}\beta_{0}}{8\pi}\ln\left(  \vec{q}_{1}^{\;2}\vec{q}_{2}%
^{\;2}\right)  \delta(\vec{q}_{11^{\prime}})\delta(\vec{q}_{22^{\prime}%
})-\left(  \psi(1)+\ln2\right)  \langle\vec{q}_{1},\vec{q}_{2}|{\hat{K}}%
^{(B)}|\vec{q}_{1}^{\,\,\prime},\vec{q}_{2}^{\,\,\prime}\rangle~.
\label{U momentum}%
\end{equation}
Note that, except the term with ${\hat{\mathcal{K}}}^{(B)}$ (which can be
omitted), it does not have the virtual part at all.

\section{ M\"{o}bius representation for the operator $\hat{U}$}
\label{sec:mobius}

We have restored the complete operator $\hat{U}_{2}$ from
its M\"{o}bius form and have obtained the total operator $\hat{U}$ in the
momentum space. But sometimes the M\"{o}bius representation can be more
convenient than the complete one. Therefore in this Section we will construct
the M\"{o}bius representation for $\hat{U}$.
Since $\hat{U}_{2}$ was originally written in
this representation~(\ref{U2}), we have to find the M\"{o}bius form for $\hat
{U}_{1}$. As was already mentioned, $\hat{U}_{1}$ has the same properties
(gauge invariance and absence of terms proportional to $\delta(\vec{q}_{1})$
or $\delta(\vec{q}_{2})$) as $\hat{\mathcal{K}}$. According to the
prescription~(\ref{K M impact}), first we need to find
\[
\langle\vec{r}_{1},\vec{r}_{2}|\alpha_{s}\hat{U}_{1}|\vec{r}_{1}^{\;\prime
},\vec{r}_{2}^{\;\prime}\rangle=\frac{\alpha_{s}N_{c}}{4\pi^{2}}\int
\frac{d\vec{q}_{1}}{2\pi}\frac{d\vec{q}_{2}}{2\pi}\frac{d\vec{k}}{(2\pi)^{2}%
}e^{i\vec{q}_{1}\vec{r}_{11^{\prime}}+i\vec{q}_{2}\vec{r}_{22^{\prime}}%
+i\vec{k}\vec{r}_{1^{\prime}2^{\prime}}}\Biggl[-\frac{2}{\vec{k}^{\,\,2}}%
\ln\vec{k}^{\,\,2}%
\]%
\[
+2\left(  \frac{\vec{k}\vec{q}_{1}}{\vec{k}^{\,\,2}\vec{q}_{1}^{\,\,2}}%
-\frac{\vec{k}\vec{q}_{2}}{\vec{k}^{\,\,2}\vec{q}_{2}^{\,\,2}}+\frac{\vec
{q}_{1}\vec{q}_{2}}{\vec{q}_{1}^{\,\,2}\vec{q}_{2}^{\,\,2}}\right)  \ln\vec
{k}^{\,\,2}+\delta(\vec{k})\left(  -\frac{\pi\beta_{0}}{2N_{c}}\ln\left(
{\vec{q}}_{1}^{\,2}{\vec{q}}_{2}^{\,2}\right)  \right.
\]%
\begin{equation}
\left.  +\int d^{2}l\left(  \frac{2}{\vec{l}^{\,\,2}}-\frac{\vec{l}(\vec
{l}-\vec{q}_{1})}{\vec{l}^{\,\,2}(\vec{l}-\vec{q}_{1})^{2}}-\frac{\vec{l}%
(\vec{l}-\vec{q}_{2})}{\vec{l}^{\,\,2}(\vec{l}-\vec{q}_{2})^{2}}\right)
\ln\vec{l}^{\,\,2}\right)  \Biggr]\;. \label{U 1 transformation}%
\end{equation}
Since in the integrand (in the square brackets) there are no terms independent
of $\vec{k}$, after the integration there will be no terms proportional to
$\delta(\vec{r}_{1^{\prime}2^{\prime}})$ which should be omitted. In
principle, the next steps are defined in the prescription~(\ref{K M impact}). But as in the
preceding Section, one must face the technical problems of separation of real
and virtual parts, since they are separately singular. In Eq.~(\ref{U 1
transformation}) the first term in the square brackets and the first
term in the integral over $\vec{l}$ are separately infrared singular and must
be treated together. These and only these terms give a contribution
proportional to $\delta(\vec{r}_{11^{\prime}})\delta(\vec{r}_{22^{\prime}})$
in the impact parameter space (and can be called ``virtual'' in this space).
But the second of them is not only infrared, but also ultraviolet divergent, so
that the coefficient of $\delta(\vec{r}_{11^{\prime}})\delta(\vec
{r}_{22^{\prime}})$ contains an ultraviolet singularity. Therefore, it must be
written in an integral form using the same trick as before:
\[
\int\frac{d\vec{q}_{1}}{2\pi}\frac{d\vec{q}_{2}}{2\pi}\frac{d\vec{k}}%
{(2\pi)^{2}}e^{i\vec{q}_{1}\vec{r}_{11^{\prime}}+i\vec{q}_{2}\vec
{r}_{22^{\prime}}+i\vec{k}\vec{r}_{1^{\prime}2^{\prime}}}\Biggl(-\frac{2}%
{\vec{k}^{\,\,2}}\ln\vec{k}^{\,\,2}+\delta(\vec{k})\int d^{2}l\frac{2}{\vec
{l}^{\,\,2}}\ln\vec{l}^{\,\,2}\Biggr)
\]%
\[
=-\delta(\vec{r}_{11^{\prime}})\delta(\vec{r}_{22^{\prime}})\int\frac{d\vec
{k}}{\vec{k}^{\,\,2}}\ln\vec{k}^{\,\,2}\left(  2e^{i\vec{k}\vec{r}_{1^{\prime
}2^{\prime}}}-2\right)  =\delta(\vec{r}_{11^{\prime}})\delta(\vec
{r}_{22^{\prime}})\int d\vec{r}_{0}\frac{d\vec{k}_{1}}{2\pi}\frac{d\vec{k}%
_{2}}{2\pi}%
\]%
\[
\times\frac{\vec{k}_{1}\vec{k}_{2}}{\vec{k}_{1}^{\;2}\vec{k}_{2}^{\;2}%
}\Biggl(e^{i\vec{k}_{1}\vec{r}_{01}+i\vec{k}_{2}\vec{r}_{02}}\ln(\vec{k}%
_{1}^{\;2}\vec{k}_{2}^{\;2})-e^{i(\vec{k}_{1}+\vec{k}_{2})\vec{r}_{01}}%
\ln(\vec{k}_{1}^{\;2})-e^{i(\vec{k}_{1}+\vec{k}_{2})\vec{r}_{02}}\ln(\vec
{k}_{2}^{\;2})\Biggr)
\]%
\[
=\delta(\vec{r}_{11^{\prime}})\delta(\vec{r}_{22^{\prime}})\int d\vec{r}%
_{0}\Biggl[\frac{\vec{r}_{12}^{\;2}}{\vec{r}_{01}^{\;2}\vec{r}_{02}^{\;2}%
}\left(  2\psi(1)+\ln4\right)
\]%
\begin{equation}
+\frac{\vec{r}_{01}\vec{r}_{02}}{\vec{r}_{01}^{\;2}\vec{r}_{02}^{\;2}}%
\ln\left(  {\vec{r}_{01}^{\;2}\vec{r}_{02}^{\;2}}\right)  -\frac{1}{\vec
{r}_{01}^{\;2}}\ln\left(  {\vec{r}_{01}^{\;2}}\right)  -\frac{1}{\vec{r}%
_{02}^{\;2}}\ln\left(  {\vec{r}_{02}^{\;2}}\right)  \Biggr]\;,
\label{infrared to ultraviolet}%
\end{equation}
where we used the integrals~(\ref{int k/k^2 ln k/k^2}). Evidently,
the representation~(\ref{infrared to ultraviolet}) is not unique. Using the
equality
\begin{equation}
\int d\vec{r}_{0}\Bigl[\frac{\vec{r}_{12}^{\;2}}{\vec{r}_{01}^{\;2}\vec
{r}_{02}^{\;2}}\ln\left(  \frac{\vec{r}_{01}^{\;2}\vec{r}_{02}^{\;2}}{(\vec
{r}_{12}^{\;2})^{2}}\right)  -\left(  \frac{1}{\vec{r}_{01}^{\;2}}-\frac
{1}{\vec{r}_{02}^{\;2}}\right)  \ln\left(  \frac{\vec{r}_{01}^{\;2}}{\vec
{r}_{02}^{\;2}}\right)  \Bigr]=0\;,
\end{equation}
we come to the representation
\[
\int\frac{d\vec{q}_{1}}{2\pi}\frac{d\vec{q}_{2}}{2\pi}\frac{d\vec{k}}%
{(2\pi)^{2}}e^{i\vec{q}_{1}\vec{r}_{11^{\prime}}+i\vec{q}_{2}\vec
{r}_{22^{\prime}}+i\vec{k}\vec{r}_{1^{\prime}2^{\prime}}}\Biggl(-\frac{2}%
{\vec{k}^{\,\,2}}\ln\vec{k}^{\,\,2}+\delta(\vec{k})\int d^{2}l\frac{2}{\vec
{l}^{\,\,2}}\ln\vec{l}^{\,\,2}\Biggr)
\]%
\begin{equation}
=\delta(\vec{r}_{11^{\prime}})\delta(\vec{r}_{22^{\prime}})\int d\vec{r}%
_{0}\frac{\vec{r}_{12}^{\;2}}{\vec{r}_{01}^{\;2}\vec{r}_{02}^{\;2}%
}\Bigl[\left(  2\psi(1)+\ln4\right)  +\ln\left(  \frac{\vec{r}_{12}^{\;2}%
}{\vec{r}_{01}^{\;2}\vec{r}_{02}^{\;2}}\right)  \Bigr]\;.
\end{equation}
The ultraviolet divergence in this (virtual in impact parameter space)
contribution must cancel analogous divergences of the other terms. Their
calculation does not require any trick. Using the integrals~(\ref{int k/k^2
ln k/k^2}), we obtain
\[
\int\frac{d\vec{q}_{1}}{2\pi}\frac{d\vec{q}_{2}}{2\pi}\frac{d\vec{k}}%
{(2\pi)^{2}}e^{i\vec{q}_{1}\vec{r}_{11^{\prime}}+i\vec{q}_{2}\vec
{r}_{22^{\prime}}+i\vec{k}\vec{r}_{1^{\prime}2^{\prime}}}\Biggl[\frac{2\vec
{q}_{1}\vec{k}}{\vec{q}_{1}^{\;2}\vec{k}^{\;2}}\ln\vec{k}^{\,\,2}+\delta
(\vec{k})\int d^{2}l\frac{(\vec{q}_{1}-\vec{l})\vec{l}}{(\vec{q}_{1}-\vec
{l})^{2}\vec{l}^{\,\,2}}\ln\vec{l}^{\,\,2}\Biggr]
\]%
\[
=\delta(\vec{r}_{22^{\prime}})\Biggl[-\frac{\vec{r}_{12}^{\;2}}{\vec
{r}_{11^{\prime}}^{\;2}\vec{r}_{21^{\prime}}^{\;2}}\Bigl(2\psi(1)+\ln
4-\ln\left(  {\vec{r}_{21^{\prime}}^{\;2}}\right)  \Bigr)+\frac{1}{\vec
{r}_{11^{\prime}}^{\;2}}\ln\left(  \frac{\vec{r}_{11^{\prime}}^{\;2}}{\vec
{r}_{21^{\prime}}^{\;2}}\right)
\]%
\begin{equation}
+\frac{1}{\vec{r}_{21^{\prime}}^{\;2}}\left(  2\psi(1)+\ln4-\ln(\vec
{r}_{21^{\prime}}^{\;2})\right)  \Biggr]\;.
\label{U 1 virtual}
\end{equation}
As follows from the representation~(\ref{K M impact}), terms in the
last line do not contribute to the M\"{o}bius representation and can
be omitted. The terms in Eq.~(\ref{U 1 transformation})
corresponding to those in the square brackets at the L.H.S. of
Eq.~(\ref{U 1 virtual}) after the substitution
$\vec{q}_{1}\leftrightarrow\vec{q}_{2},\vec{k}\leftrightarrow-\vec{k}$
give a contribution equal to the R.H.S. of Eq.~(\ref{U 1 virtual})
after the substitution $\vec
{r}_{1}\leftrightarrow\vec{r}_{2},\vec{r}_{1}^{\;\prime}\leftrightarrow\vec
{r}_{2}^{\;\prime}$. To calculate the remaining terms of~(\ref{U 1
transformation}), one needs, besides the integrals~(\ref{int k/k^2
ln k/k^2}), only the Fourier transform of $\ln\vec{q}^{\;2}$. Since
it is singular, it requires regularization (i.e. extension of the
definition). It was  considered in detail in Section
\ref{sec:interrelation}. Choosing the functions $f_1$ and $f_2$ in
Eq.~(\ref{K M impact}) from consideration of simplicity, we obtain
\[
\langle\vec{r}_{1}\vec{r}_{2}|\alpha_{s}\hat{U}_{1M}|\vec{r}_{1}^{\;\prime
}\vec{r}_{2}^{\;\prime}\rangle\!=\!\frac{\alpha_{s}N_{c}}{4\pi^{2}}\int d\vec
{r}_{0}\Biggl\{\delta(\vec{r}_{11^{\prime}})\delta(\vec{r}_{02^{\prime}%
})\left[  \frac{\vec{r}_{12}^{\,\,2}\ln\left(  \vec{r}_{01}^{\,\,2}\right)
}{\vec{r}_{01}^{\,\,2}\vec{r}_{02}^{\,\,2}}+\frac{1}{\vec{r}_{02}^{\,\,2}}%
\ln\left(  \frac{\vec{r}_{02}^{\,\,2}}{\vec{r}_{01}^{\,\,2}}\right)  \right]
\]%
\[
+\delta(\vec{r}_{22^{\prime}})\delta(\vec{r}_{01^{\prime}})\left[  \frac
{\vec{r}_{12}^{\,\,2}\ln\left(  \vec{r}_{02}^{\,\,2}\right)  }{\vec{r}%
_{01}^{\,\,2}\vec{r}_{02}^{\,\,2}}+\frac{1}{\vec{r}_{01}^{\,\,2}}\ln\left(
\frac{\vec{r}_{01}^{\,\,2}}{\vec{r}_{02}^{\,\,2}}\right)  \right]
+\delta(\vec{r}_{11^{\prime}})\delta({r}_{22^{\prime}})\frac{\vec{r}_{12}%
^{\,\,2}\ln\left(  \frac{\vec{r}_{12}^{\,\,2}}{\vec{r}_{01}^{\,\,2}\vec{r}%
_{02}^{\,\,2}}\right)  }{\vec{r}_{01}^{\,\,2}\vec{r}_{02}^{\,\,2}}\Biggr\}
\]%
\[
+\frac{1}{\pi\vec{r}_{1^{\prime}2^{\prime}}^{\,\,2}}\left[  \frac{2\vec
{r}_{11^{\prime}}\vec{r}_{22^{\prime}}}{\vec{r}_{11^{\prime}}^{\,\,2}\vec
{r}_{22^{\prime}}^{\,\,2}}-\frac{\vec{r}_{11^{\prime}}\vec{r}_{12^{\prime}}%
}{\vec{r}_{11^{\prime}}^{\,\,2}\vec{r}_{12^{\prime}}^{\,\,2}}-\frac{\vec
{r}_{21^{\prime}}\vec{r}_{22^{\prime}}}{\vec{r}_{21^{\prime}}^{\,\,2}\vec
{r}_{22^{\prime}}^{\,\,2}}\right]  -\left(  \psi(1)+\ln2\right)  \langle
\vec{r}_{1},\vec{r}_{2}|{\hat{K}}_{M}^{(B)}|\vec{r}_{1}^{\,\,\prime},\vec{r}%
_{2}^{\,\,\prime}\rangle~
\]%
\begin{equation}
+\frac{\alpha_{s}\beta_{0}}{8\pi^{2}}\Biggl[\delta(\vec{r}_{11^{\prime}%
})\left(  \frac{1}{(\vec{r}_{22^{\prime}}^{\,\,2})_{R}}-\frac{1}{\vec
{r}_{12^{\prime}}^{\,\,2}}\right)  +\delta(\vec{r}_{22^{\prime}})\left(
\frac{1}{(\vec{r}_{11^{\prime}}^{\,\,2})_{R}}-\frac{1}{\vec{r}_{21^{\prime}%
}^{\,\,2}}\right)  \Biggr]~, \label{U1 coordinate}%
\end{equation}
where $(1/\vec{r}^{\;2})_R$ is defined as in (\ref{int xs r}), (\ref{definition xs r}) and ${\hat{K}}_{M}^{(B)}$ is the leading order BFKL kernel in the M\"{o}bius
representation
\[
\langle\vec{r}_{1}\vec{r}_{2}|{\hat{\mathcal{K}}}_{M}^{(B)}|\vec{r}%
_{1}^{\;\prime}\vec{r}_{2}^{\;\prime}\rangle=\frac{\alpha_{s}N_{c}}{2\pi^{2}%
}\int d\vec{r}_{0}\frac{\vec{r}_{12}^{\,\,2}}{\vec{r}_{01}^{\,\,2}\vec{r}%
_{02}^{\,\,2}}%
\]%
\begin{equation}
\times\Biggl[\delta(\vec{r}_{11^{\prime}})\delta(\vec{r}_{02^{\prime}}%
)+\delta(\vec{r}_{01^{\prime}})\delta(\vec{r}_{22^{\prime}})-\delta(\vec
{r}_{11^{\prime}})\delta({r}_{22^{\prime}})\Biggr]~. \label{K B M}%
\end{equation}
The M\"{o}bius form of the total operator ${\hat{U}}$ is given by the sum
of the two pieces expressed in Eqs.~(\ref{U1 coordinate}) and~(\ref{U2}):
\[
\langle\vec{r}_{1}\vec{r}_{2}|\alpha_{s}\hat{U}_{M}|\vec{r}_{1}^{\;\prime}%
\vec{r}_{2}^{\;\prime}\rangle=\frac{\alpha_{s}N_{c}}{4\pi^{2}}\int d\vec
{r}_{0}\Biggl\{\delta(\vec{r}_{11^{\prime}})\delta(\vec{r}_{02^{\prime}%
})\left[  \frac{\vec{r}_{12}^{\,\,2}}{\vec{r}_{01}^{\,\,2}\vec{r}_{02}^{\,\,2}%
}\ln\left(  \frac{\vec{r}_{12}^{\,\,2}}{\vec{r}_{02}^{\,\,2}}\right)  \right.
\]%
\[
\left.  +\frac{1}{\vec{r}_{02}^{\,\,2}}\ln\left(  \frac{\vec{r}_{02}^{\,\,2}%
}{\vec{r}_{01}^{\,\,2}}\right)  \right]  +\delta(\vec{r}_{22^{\prime}}%
)\delta(\vec{r}_{01^{\prime}})\left[  \frac{\vec{r}_{12}^{\,\,2}}{\vec{r}%
_{01}^{\,\,2}\vec{r}_{02}^{\,\,2}}\ln\left(  \frac{\vec{r}_{12}^{\,\,2}}%
{\vec{r}_{01}^{\,\,2}}\right)  +\frac{1}{\vec{r}_{01}^{\,\,2}}\ln\left(
\frac{\vec{r}_{01}^{\,\,2}}{\vec{r}_{02}^{\,\,2}}\right)  \right]  \Biggr\}
\]%
\[
+\frac{1}{\pi\vec{r}_{1^{\prime}2^{\prime}}^{\,\,2}}\left[  \frac{2\vec
{r}_{11^{\prime}}\vec{r}_{22^{\prime}}}{\vec{r}_{11^{\prime}}^{\,\,2}\vec
{r}_{22^{\prime}}^{\,\,2}}-\frac{\vec{r}_{11^{\prime}}\vec{r}_{12^{\prime}}%
}{\vec{r}_{11^{\prime}}^{\,\,2}\vec{r}_{12^{\prime}}^{\,\,2}}-\frac{\vec
{r}_{21^{\prime}}\vec{r}_{22^{\prime}}}{\vec{r}_{21^{\prime}}^{\,\,2}\vec
{r}_{22^{\prime}}^{\,\,2}}\right]  -\left(  \psi(1)+\ln2\right)  \langle
\vec{r}_{1},\vec{r}_{2}|{\hat{K}}_{M}^{B}|\vec{r}_{1}^{\,\,\prime},\vec{r}%
_{2}^{\,\,\prime}\rangle~
\]%
\begin{equation}
+\frac{\alpha_{s}\beta_{0}}{8\pi^{2}}\Biggl[\delta(\vec{r}_{11^{\prime}%
})\left(  \frac{1}{(\vec{r}_{22^{\prime}}^{\,\,2})_{R}}-\frac{1}{\vec
{r}_{12^{\prime}}^{\,\,2}}\right)  +\delta(\vec{r}_{22^{\prime}})\left(  \frac
{1}{(\vec{r}_{11^{\prime}}^{\,\,2})_{R}}-\frac{1}{\vec{r}_{21^{\prime}%
}^{\,\,2}}\right)  \Biggr]~. \label{U coordinate}%
\end{equation}
It is worthwhile to mention that transferring Eq.~(\ref{U momentum}) into
the M\"{o}bius representation in the impact parameter space in accordance with
the prescription~(\ref{K M impact}) gives exactly the 
result~(\ref{U coordinate}).

\section{Conclusion}
\label{sec:conclusion}

We investigated the connection between the complete and M\"{o}bius
representations of gauge invariant operators, taking particular
care of the BFKL kernel for scattering of colourless particles.
Following Ref.~\cite{Bartels:2004ef} we call M\"{o}bius representation (form) of
some two-particle (or two-Reggeon, as in the case of BFKL kernel) operator its
form in the space of functions vanishing at coinciding impact parameters of
these particles. In this representation the BFKL kernel has remarkable
properties.
In the leading order it is invariant with respect to the group of
M\"{o}bius transformations of impact parameters~\cite{Lipatov:1985uk}, and in
the NLO it can be transformed into a simple quasi-conformal shape. An
important question is the possibility of restoration of the complete kernel
from this shape. It is evident that a generic operator cannot be completely 
restored from its M\"obius representation, since in this representation it
acts in a truncated space of functions. Moreover, the explicit form
of a generic operator in this representation can be written only
in the coordinate space. The transformation into the momentum space can
be impossible because of the singularity at coinciding impact parameters.
However, this is not the case for the BFKL kernel, for which M\"obius and
complete representations are equivalent. The reason for that is the gauge
invariance of the kernel.

In this paper it has been shown that for any gauge invariant
two-particle operator it is possible to restore the complete
operator from its M\"{o}bius representation. We have shown that the
restoration is unique up to terms proportional to
$\delta(\vec{q}_{1})$ or $\delta(\vec{q}_{2})$ and symmetry with
respect to the Reggeon exchange. It  was  explicitly demonstrated
for the operator responsible for the transformation of the standard
BFKL kernel to the quasi-conformal shape. Originally this operator
was presented as a sum of two pieces, one of them was found acting
in the complete representation in the momentum space and the other
in the M\"{o}bius representation in the coordinate space. We found
both M\"{o}bius and complete representations of the full operator.

\vspace{0.5cm} {\textbf{{\Large Acknowledgments}}}

\vspace{0.5cm} V.S.F. thanks the Dipartimento di Fisica
dell'Universit\`{a} della Calabria and the Istituto Nazionale di
Fisica Nucleare (INFN), Gruppo Collegato di Cosenza, for warm hospitality
while part of this work was done and for financial support. He
thanks also the Galileo Galilei Institute for Theoretical Physics
for the hospitality and the INFN for the partial support during the
completion of this work.

\end{document}